\documentclass[12pt]{article}

\usepackage{amssymb}
\usepackage{amsmath}
\usepackage{amscd}
\usepackage{latexsym}

\usepackage{graphicx}
\usepackage{array}

\usepackage{cite}

\newcommand{\be}{\begin{equation}}
\newcommand{\ee}{\end{equation}}

\newcommand{\dlt}{\delta}
\newcommand{\prt}{\partial}
\newcommand{\br}{{\bf r}}
\newcommand{\bk}{{\bf k}}

\newcommand{\bt}{\beta}
\newcommand{\vp}{\varphi}
\newcommand{\ep}{\varepsilon}

\newcommand{\ra}{\rightarrow}
\newcommand{\sgm}{\sigma}

\newcommand{\gm}{\gamma}
\newcommand{\om}{\omega}

\newcommand{\dgr}{\dagger}
\newcommand{\lbd}{\lambda}
\newcommand{\Lbd}{\Lambda}

\newcommand{\rgl}{\rangle}
\newcommand{\lgl}{\langle}

\begin{document}

\begin{center}
{\Large{\bf Local condensate depletion at trap center under strong interactions} \\ [5mm]

V.I.~Yukalov$^{1,2,*}$ and E.P.~Yukalova$^{3}$ } \\ [3mm] 

{\it 
$^1$Bogolubov Laboratory of Theoretical Physics, \\
Joint Institute for Nuclear Research, Dubna 141980, Russia \\ [2mm]
                                           
$^2$Instituto de Fisica de S\~ao Calros, Universidade de S\~ao Paulo, \\
CP 369,  S\~ao Carlos 13560-970, S\~ao Paulo, Brazil  \\ [2mm]

$^3$Laboratory of Information Technologies, \\
Joint Institute for Nuclear Research, Dubna 141980, Russia }

\end{center}

\vskip 1cm

\begin{abstract}
Cold trapped Bose-condensed atoms, interacting via hard-sphere repulsive potentials 
are considered. Simple mean-field approximations show that the condensate distribution 
inside a harmonic trap always has the shape of a hump with the maximum condensate 
density occurring at the trap center. However Monte Carlo simulations at high density 
and strong interactions display the condensate depletion at the trap center. The 
explanation of this effect of local condensate depletion at trap center is suggested in the 
frame of self-consistent theory of Bose-condensed systems. The depletion is shown to 
be due to the existence of the anomalous average that  takes into account pair correlations 
and appears in systems with broken gauge symmetry.    

\end{abstract}

\vskip 1cm

{\parindent =0pt


\vskip 1cm

$^*${\bf Corresponding author}: V.I. Yukalov

\vskip 2mm

{\bf E-mail}: yukalov@theor.jinr.ru     }

\newpage

\section{Introduction}

Bose-Einstein condensed systems of cold trapped atoms with hard-core repulsive 
interactions are usually studied for small gas parameter $\gm\equiv\rho^{1/3}a_s$, 
where $\rho$ is average density and $a_s$ is $s$-wave 
scattering length (hard-core diameter) 
(see, e.g., \cite{Pethick_1,Bloch_2,Yurovsky_3,Krems_4,Cote_5}.
This assumes low average density and weak interactions. The standard Thomas-Fermi 
approximation \cite{Pethick_1}, gives, under these conditions, hump-shaped equilibrium 
spatial distribution of condensed atoms inside a trap, with the maximal local density 
$\rho_0(0)$  at the trap center ${\bf r}= 0$. The Bogolubov theory \cite{Bogolubov_6}, 
describing experimentally observable condensate depletion due to higher order momentum 
states \cite{Xu_7}, also shows that the condensate profile exhibits a maximum at the trap 
center \cite{Javanainen_48}.

Moreover, the Thomas-Fermi approximation is known to be asymptotically exact for large 
coupling parameters, which assumes that the same situation, with the maximal condensate 
density at the trap center, should remain for arbitrarily strong interactions. This also concerns 
other, more refined, solutions of the nonlinear Schrodinger (NLS) equation, such as the 
optimized variational approximation \cite{Yukalov_2}. A very accurate solution of the NLS 
equation is achieved by employing self-similar approximants \cite{Yukalov_3}, which alow 
for obtaining accurate solutions in the whole range of the coupling parameter, correctly 
interpolating between weak-coupling and strong-coupling asymptotic forms \cite{Yukalov_4}. 
In all these cases, the ground-state condensate density distribution reaches a maximum 
at the trap center.     
  
Contrary to these solutions, Monte Carlo simulations \cite{Dubois_5,Dubois_6} 
demonstrate that, although at small gas parameters the condensate density maximum 
is really at the trap center, but starting from the gas parameter $\gm\approx 0.3$ 
the condensate at the trap center becomes depleted, so that at the trap center there
appears a local minimum of the condensate density.  This result also was obtained 
in the slave-boson approach \cite{Ziegler_7}.  

Since, as is emphasized above, different solutions of the NLS equation always lead 
to the condensate fraction distribution with a maximum at the trap center 
\cite{Proukakis_8,Yukalov_9}, there exists the general belief that the local condensate 
depletion at the trap center cannot be explained by any mean-field approximation. 

The aim of the present paper is to demonstrate that the effect of local condensate
depletion at trap center can be straightforwardly explained in the frame of self-consistent
approach even on the level of Thomas-Fermi approximation. Let us emphasize that
we consider this effect of trap-center condensate depletion for atoms interacting through
isotropic short-range potentials, such as contact potentials that model the hard-core 
potentials. The situation for long-range dipolar condensates is rather different due to the 
anisotropy of interactions \cite{Griesmaier_10,Baranov_11,Baranov_12,Gadway_13}, 
which is not touched here.  

Throughout the paper, we use the system of units with the Planck and Boltzmann constants 
set to unity.

\section{Main equations}

Here we briefly mention the basic points of the self-consistent approach 
\cite{Yukalov_15,Yukalov_16,Yukalov_17} to be used below. 

We consider the contact interaction potential 
\be
\label{1}
 \Phi(\br) = \Phi_0\dlt(\br) \; , \qquad \Phi_0 = 4\pi \; \frac{a_s}{m} \;  ,
\ee
in which $a_s > 0$ is $s$-wave scattering length and $m$ is mass. The energy Hamiltonian
for a system of $N$ Bose atoms reads as
\be
\label{2}
 \hat H = \int \hat\psi^\dgr(\br) \left [ - \; \frac{\nabla^2}{2m} + U(\br) 
\right ] \hat\psi(\br) \; d\br  +  \frac{1}{2} \; \Phi_0
\int \hat\psi^\dgr(\br) \hat\psi^\dgr(\br) \hat\psi(\br) \hat\psi(\br)\; d\br \; ,
\ee
where $U(\bf r)$ is a trapping potential. Time dependence of the field operators, for
the compactness of notation, is suppressed.  

Under Bose-Einstein condensation, the global gauge symmetry of the system becomes
broken, which is the necessary and sufficient condition for Bose condensation
\cite{Lieb_18,Yukalov_19,Yukalov_20}. The simplest way of gauge symmetry breaking 
can be done by means of the Bogolubov shift \cite{Bogolubov_21,Bogolubov_22,Bogolubov_23}
\be
\label{3}
 \hat\psi(\br) = \eta(\br) + \psi_1(\br) \;  ,
\ee
where $\eta(\bf r)$ is the condensate function and $\psi_1(\bf r)$ is the field operator
of uncondensed atoms, satisfying the Bose commutation relations. It is worth stressing
that equation (\ref{3}) is an exact canonical transformation \cite{Yukalov_24}, but not an 
approximation, as one sometimes writes. The functional variables of condensed and 
uncondensed atoms are mutually orthogonal. The condensate function plays the role
of a functional order parameter, which implies for the statistical averages the equalities 
\be 
\label{4}
 \eta(\br) \equiv \lgl \hat\psi(\br) \rgl \; , \qquad \lgl \psi_1(\br) \rgl = 0 \;  .
\ee
The condensate function is normalized to the number of condensed atoms
\be
\label{5}
 N_0 = \int |\eta(\br) |^2 \; d\br \;  ,
\ee
while the number of uncondensed atoms is given by the statistical average
\be
\label{6}
 N_1 = \lgl \hat N_1 \rgl \; , \qquad 
\hat N_1 = \int \psi_1^\dgr(\br) \psi_1(\br) \; d\br \;  ,
\ee
the total number of atoms being $N = N_0 + N_1$. 

The evolution equations are prescribed by the extremization of an effective action,
under the validity of the above conditions (\ref{4}), (\ref{5}) and (\ref{6}), which
is equivalent to the equations of motion for the condensate function
\be
\label{7}
 i\; \frac{\prt}{\prt t}  \; \eta(\br,t) = 
\left \lgl \frac{\dlt H}{\dlt\eta^*(\br,t)} \right \rgl
\ee
and for the operator of uncondensed atoms
\be
\label{8}
 i\; \frac{\prt}{\prt t}  \; \psi_1(\br,t) = 
 \frac{\dlt H}{\dlt\psi_1^\dgr(\br,t)} \;  ,
\ee
where the grand Hamiltonian
\be
\label{9}
 H = \hat H - \mu_0 N_0 - \mu_1 \hat N_1  - \hat\Lbd \;   ,
\ee
with
$$
 \hat\Lbd \equiv \int \left [ \lbd(\br) \psi_1^\dgr(\br) + 
\lbd^*(\br) \psi_1(\br) \right ] \; d\br \;  ,
$$
includes the Lagrange multipliers $\mu_0$, $\mu_1$, and $\lambda(\bf r)$ that guarantee
the validity of the required conditions (\ref{4}), (\ref{5}), and (\ref{6}). 

For an equilibrium system, the statistical operator is defined as the minimizer of the
information functional, under the given conditions, which results in the operator
\be
\label{10}
 \hat\rho = \frac{1}{Z} \; e^{-\bt H} \; , \qquad Z \equiv {\rm Tr} e^{-\bt H} \;  ,
\ee
with the same grand Hamiltonian (\ref{9}) and $\beta \equiv 1/T$ being the inverse
temperature. The statistical operator (\ref{10}), taking into account the conditions 
uniquely representing the system, together with the Fock space ${\cal F}(\psi_1)$, 
generated by the field operator $\psi_1$, forms the representative statistical 
ensemble $\{\mathcal{F}(\psi_1),\hat\rho\}$. More details can be found in the review 
articles \cite{Yukalov_9,Yukalov_25}. 

It is important to stress that the self-consistent approach, delineated above, respects 
all conservation laws and thermodynamic relations, at the same time yielding a gapless 
spectrum of collective excitations, in agreement with the theorems by Bogolubov 
\cite{Bogolubov_21,Bogolubov_22} and Hugenholtz and Pines \cite{Hugenholtz_26},      
thus, resolving the Hohenberg-Martin \cite{Hohenberg_27} dilemma of conserving 
versus gapless theories.

\section{Equilibrium system}

In the case of an equilibrium system in a slowly varying trapping potential, a reasonably 
accurate description can be done by resorting to the local-density approximation 
\cite{Pethick_1,Yukalov_9,Yukalov_20,Purwanto_28}. Then one can treat the trapped 
atomic cloud in a way similar to the uniform case, with the spatial dependence coming 
through the local densities. Note that there exist the traps, in which the density 
distribution is almost uniform \cite{Gotlibovych_29}. 

Below, we use the local density approximation combined with the Hartree-Fock-Bogolubov 
(HFB) approximation. For homogeneous Bose-condensed systems, the HFB approximation, 
in the frame of the self-consistent approach, sketched in Sec. 2, was shown to provide 
an accurate description, being in a very good quantitative agreement with Monte Carlo 
simulations \cite{Giorgini_30,Pilati_31,Rossi_32} for Bose-Einstein condensate fraction 
and ground-state energy in the whole range of the gas parameter form zero up to its 
values corresponding to the system freezing \cite{Yukalov_33}. And the Bose 
condensation phase transition was shown \cite{Yukalov_34} to be of second order, 
as it has to be for this class of $XY$ equivalence \cite{Matsubara_35,Floerchinger_36}.  
The phase transition critical temperature in the mean-filed HFB approximation coincides 
with that of the ideal Bose gas, which is again the general feature of mean-field 
approximations. To find the critical temperature shift, caused by atomic interactions, 
one needs to go above the mean-field approach, as is discussed in the review articles
\cite{Andersen_37,Yukalov_38,Yukalov_39}.  
  
In numerical Monte Carlo simulations, one employs the hard-core interaction potential
\cite{Giorgini_30,Pilati_31,Rossi_32}. In theoretical calculations, the divergent hard-core
potential is treated by means of t-matrix approximation \cite{Kim_40,Kim_41} or by the 
method of smoothed potentials \cite{Yukalov_42}. The hard-core interaction potential
is known to well represent more realistic interaction potentials, such as the Lennard-Jones 
potential, even for rather dense fluids, e.g. for liquid helium \cite{Kalos_43}. At small 
values of the gas parameter, the system properties have been shown to be universal, 
being almost independent of the particular shapes of interaction potentials, which makes 
it possible to use the contact potential \cite{Giorgini_30}. Moreover, it has also been 
demonstrated \cite{Yukalov_33} that in the frame of the self-consistent HFB approximation 
the contact potential (\ref{1}) well represents the results, obtained for the hard-core potential,
even for sufficiently large gas parameters. 

Thus, below we employ the local-density HFB approximation, with the interaction potential   
(\ref{1}).  

The density of condensed atoms is
\be
\label{11}
 \rho_0(\br) \equiv | \eta(\br) |^2 \;  .
\ee
The density of uncondensed atoms is given by 
\be
\label{12}
 \rho_1(\br) \equiv \lgl \psi_1^\dgr(\br) \psi_1(\br) \rgl \;  .
\ee
And the anomalous average is denoted as
\be
\label{13}
 \sgm_1(\br) \equiv \lgl \psi_1(\br) \psi_1(\br) \rgl \;   .
\ee

The condensate function equation (\ref{7}), for an equilibrium system, takes 
the form
\be
\label{14}
 \hat H(\br) \eta(\br) = \mu_0 \eta(\br) \;  ,
\ee
where
\be
\label{15}
 \hat H(\br) = -\; \frac{\nabla^2}{2m} + U(\br) +  \Phi_0 [ \rho_0(\br) +
2\rho_1(\br) + \sgm_1(\br) ] \;  .
\ee

The density of uncondensed atoms and the anomalous average (\ref{13}) can be 
represented through the integrals 
\be
\label{16}
 \rho_1(\br) = \int n_k(\br) \; \frac{d\bk}{(2\pi)^3} \;  , \qquad
\sgm_1(\br) = \int \sgm_k(\br) \; \frac{d\bk}{(2\pi)^3} \;
\ee
in which the local momentum distribution of uncondensed atoms is
\be
\label{17}
n_k(\br) = \frac{\om_k(\br)}{2\ep_k(\br)} \; 
{\rm coth} \left ( \frac{\ep_k(\br)}{2T} \right ) - \; \frac{1}{2}
\ee
and the local anomalous average, depending on momentum, is
\be
\label{18}
\sgm_k(\br) = -\; \frac{mc^2(\br)}{2\ep_k(\br)} \; 
{\rm coth} \left ( \frac{\ep_k(\br)}{2T} \right ) \;  .
\ee
Here the notation 
$$
\om_k(\br) \equiv \frac{k^2}{2m} + mc^2
$$ 
is used. And $\varepsilon_k(\bf r)$ is the local spectrum of collective excitations
\be
\label{19}
 \ep_k(\br) = \sqrt{c^2(\br) k^2 + \left ( \frac{k^2}{2m}\right )^2 } \;  .
\ee
The local sound velocity $c(\bf r)$ satisfies the equation
\be
\label{20}
 mc^2(\br) = \Phi_0 [ \rho_0(\br) + \sgm_1(\br) ] \;  .
\ee
The total local density of atoms is the sum
\be
\label{21}
 \rho(\br) = \rho_0(\br) + \rho_1(\br) \;  .
\ee
Keeping in mind the case of zero temperature, the density of uncondensed atoms
reads as
\be
\label{22}
 \rho_1(\br) = \frac{m^3}{3\pi^2} \; c^3(\br) \qquad ( T = 0 ) \;  .
\ee

In the Bogolubov approximation \cite{Bogolubov_21,Bogolubov_22}, valid for
asymptotically weak interactions, the anomalous average is negligibly small
as compared to the density $\rho_0 \simeq \rho$. Then equation (\ref{20}) reduces
to the Bogolubov sound
\be
\label{23}
 c_B(\br) = \frac{1}{m} \; \sqrt{m\Phi_0\rho(\br) } \;  .
\ee

But, in general, the existence of the anomalous average is extremely important.
First of all, the appearance of the anomalous average occurs simultaneously with 
the arising Bose-Einstein condensate, both of them being caused by the global 
gauge symmetry breaking. It is possible to show \cite{Yukalov_9} that, strictly 
speaking, if $\sigma_1$ is set to zero, then the condensate fraction also becomes 
zero. If this self-consistency is broken by setting the anomalous average to zero,
while keeping a finite condensate fraction, then there appears a number of 
unreasonable consequences: the Bose condensation transition becomes of 
incorrect first order, superfluid fraction becomes negative, isothermal compressibility
diverges, meaning instability \cite{Yukalov_9,Yukalov_17,Yukalov_20,Yukalov_38}.
In addition, as we show below, the presence of the anomalous average explains
the effect of local trap center depletion under strong interactions. 

Calculating the anomalous average for atoms with contact interaction potential, one
meets the problem of divergence. However, this problem can be overcome by resorting
to some kind of regularization \cite{Olshani_44}. We find it the most convenient to 
employ dimensional regularization \cite{Yukalov_9,Yukalov_20,Andersen_37}. This 
regularization is exact under asymptotically weak interactions, and can be analytically 
continued to finite interaction strength. The related procedure, at zero temperature, 
leads \cite{Yukalov_9,Yukalov_33} to the iterative equation
\be
\label{24}
\sgm_1^{(n+1)}(\br) = \frac{m^2}{\pi^2} \; c_B^2(\br) 
\left\{ m\Phi_0 \left [ \rho_0(\br) + \sgm_1^{(n)}(\br) \right ] \right \}^{1/2}
\ee
for the averages of $n$-th order. To second order, we obtain
\be
\label{25}
 \sgm_1(\br) = \frac{m^2}{\pi^2} \; c_B^2(\br) \left\{ 
m\Phi_0 \left [ \rho_0(\br) + 
\frac{m^2}{\pi^2} \; c_B^2(\br) \; \sqrt{m\Phi_0\rho_0(\br)} 
\right ] \right \}^{1/2}  .
\ee
In the following, we use this expression for the anomalous average that can also
be rewritten as
$$
 \sgm_1(\br) = \frac{m^3}{\pi^2} \; c_B^2(\br) \; \sqrt{c_0^2(\br) +
\frac{m^2}{\pi^2} \; \Phi_0 c_B^2(\br) c_0(\br)} \; ,
$$
where
$$
 c_0(\br) \equiv \frac{1}{m} \; \sqrt{m\Phi_0\rho_0(\br)} \;  .
$$

The anomalous average (\ref{25}) enjoys the necessary conditions: It becomes zero
together with the condensate density, when gauge symmetry is not broken, and is finite
as soon as the condensate density is nonzero, when gauge symmetry is broken, thus 
satisfying the {\it symmetry breaking condition}
\be
\label{26}
 \sgm_1(\br) \ra 0 \qquad ( \rho_0 \ra 0 ) \;  .
\ee
And it tends to zero in the limit of zero interactions, when the system turns into ideal 
Bose gas, that is, obeying the {\it ideal gas condition}, 
\be
\label{27}
 \sgm_1(\br) \ra 0 \qquad ( \Phi_0 \ra 0 ) \;  .
\ee
   
The total number of atoms 
\be
\label{28}
N = \int \rho(\br) \; d\br = N_0 + N_1
\ee
is the sum of the numbers of condensed and uncondensed atoms,
\be
\label{29}
 N_0 = \int \rho_0(\br) \; d\br \; , \qquad N_1 = \int \rho_1(\br) \; d\br \;  .
\ee
The mean condensate fraction and the fraction of uncondensed atoms are
\be
\label{30}
\overline n_0 \equiv \frac{N_0}{N} = \frac{1}{N} \int \rho_0(\br) \; d\br \; ,
\qquad
\overline n_1 \equiv \frac{N_1}{N} = \frac{1}{N} \int \rho_1(\br) \; d\br 
\ee
respectively. As is clear, 
$$
 \overline n_0 + \overline n_1 = 1\;  .
$$

\section{Thomas-Fermi approximation}

In the Thomas-Fermi approximation, one neglects kinetic energy as compared to 
the potential energy of atoms \cite{Pethick_1}. This approximation is asymptotically
exact in the limit of large gas parameters and provides a reasonable description
even at rather small interactions \cite{Caracanhas_45}. 

Neglecting the kinetic term in the condensate function equation (\ref{14}) yields
\be
\label{31}
 \rho_0(\br) = \frac{\mu_0-U(\br)}{\Phi_0} \; - \; 2\rho_1(\br) - \sgm_1(\br) \;  ,
\ee
which is valid inside the Thomas-Fermi volume, where this expression is non-negative.
The boundary conditions are
\be
\label{32}
  \rho_0(\br_{TF}) = \rho_1(\br_{TF}) = \sgm_1(\br_{TF}) = 0 \;,
\ee
where ${\bf r}_{TF}$ is the Thomas-Fermi boundary vector running over the Thomas-Fermi 
surface.  From this boundary condition, it follows
\be
\label{33}
 \mu_0 = U(\br_{TF} ) \; .
\ee

Assuming a spherically symmetric harmonic trapping potential
\be
\label{34}
 U(\br) = \frac{m}{2} \; \om_0^2 | \br |^2  
\ee
gives the chemical potential
\be
\label{35}
\mu_0 = \frac{m}{2} \; \om_0^2 r_{TF}^2 \;   ,
\ee
with the Thomas-Fermi radius
\be
\label{36}
 r_{TF} \equiv | \br_{TF} | = \sqrt{ \frac{2\mu_0}{m\om_0^2} } \;  .
\ee
The ratio
$$
\frac{r_{TF}}{l_0} = \sqrt{ \frac{2\mu_0}{\om_0} } \qquad \left ( l_0 \equiv
\frac{1}{\sqrt{m\om_0} } \right )
$$
can be larger than one, since the Thomas-Fermi radius is usually larger than 
the oscillator length $l_0$ because of atomic interactions. 

It is convenient to introduce the dimensionless spherical variable
\be
\label{37}
r \equiv \frac{|\br|}{r_{TF} } \qquad ( 0 \leq r \leq 1 ) \;  .
\ee
The dimensionless fractions of condensed and uncondensed atoms are
\be
\label{38}
 n_0(r) \equiv \frac{\rho_0(\br)}{\rho} \; , \qquad  
n_1(r) \equiv \frac{\rho_1(\br)}{\rho} \; ,
\ee
where $\rho \equiv N/V$ is average atomic density. The dimensionless total density
reads as
\be
\label{39}
 n(r) \equiv \frac{\rho(\br)}{\rho} = n_0(r) + n_1(r) \;  .
\ee
Also, we define the dimensionless anomalous average
\be
\label{40}
\sgm(r) \equiv \frac{\sgm_1(\br)}{\rho} \;   .
\ee

The interaction strength is characterized by the gas parameter
\be
\label{41}
 \gm \equiv \rho^{1/3} a_s \;  .
\ee

The dimensionless sound velocity and its Bogolubov approximation are
\be
\label{42}
s(r) \equiv \frac{mc(\br)}{\rho^{1/3}}
\ee
and, respectively,
\be
\label{43}
 s_B(r) \equiv \frac{mc_B(\br)}{\rho^{1/3} } = \sqrt{4\pi\gm n(\br) } \;  .
\ee
The dimensionless chemical potential is
\be
\label{44}
 \mu \equiv \frac{m\mu_0}{\rho^{2/3}} = 
\frac{m^2\om_0^2}{2\rho^{2/3}} \; r_{TF}^2 \;  .
\ee
Then equation (\ref{31}) is reduced to the form
\be
\label{45}
 n_0(r) = \frac{\mu}{4\pi\gm} \; ( 1 - r^2 ) - 2n_1(r) - \sgm(r) \;  ,
\ee
with the local fraction of uncondensed atoms
\be
\label{46}
n_1(r) = \frac{s^3(r)}{3\pi^2} \; ,
\ee
local anomalous average
\be
\label{47}
 \sgm(r) = \frac{s_B^2(r)}{\pi^2} \left \{ 4\pi\gm \left [ n_0(r) + 
\frac{s_B^2(r)}{\pi^2}\; \sqrt{4\pi\gm n_0(r)} \right ] \right \}^{1/2} \;  ,
\ee
and the sound velocity defined by the equation
\be
\label{48}
 s^2(r) = 4\pi\gm [ n_0 + \sgm(r) ] \;  .
\ee
In view of notation (\ref{43}), the anomalous average takes the form
\be
\label{49}
 \sgm(r) = \frac{8}{\sqrt{\pi}} \; \gm^{3/2} n(r) \left [ n_0(r) + 
\frac{8}{\sqrt{\pi} }\; \gm^{3/2} n(r) \sqrt{n_0(r) } \right ]^{1/2}  .
\ee

The mean fractions of condensed and uncondensed atoms can be represented as
\be
\label{50}
 \overline n_0 = \frac{4\pi}{N} \; \rho r_{TF}^3 \int_0^1 n_0(r) r^2 \; dr \; , 
\qquad 
\overline n_1 = \frac{4\pi}{N} \; \rho r_{TF}^3 \int_0^1 n_1(r) r^2 \; dr \; .
\ee
The normalization condition (\ref{28}) yields
\be
\label{51}
\frac{4\pi}{N} \; \rho r_{TF}^3 \int_0^1 n(r) r^2 \; dr = 1 \;   .
\ee

The atomic cloud occupies the Thomas-Fermi volume $V = V_{TF}$, so that the mean 
density reads as
\be
\label{52}
 \rho = \frac{N}{V_{TF} } \; , \qquad 
V_{TF} = \frac{4\pi}{3} \; r_{TF}^3 \;  .
\ee
Therefore the mean fractions (\ref{50}) are defined by the expressions
\be
\label{53}
\overline n_0 = 3 \int_0^1 n_0(r) r^2 \; dr \; , \qquad
\overline n_1 = 3 \int_0^1 n_1(r) r^2 \; dr \;  .
\ee
And the normalization condition (\ref{51}) becomes
\be
\label{54}
 3 \int_0^1 n(r) r^2 \; dr = 1 \; ,
\ee
which defines the chemical potential
\be
\label{55}
 \mu = \frac{1}{2} \; \left ( \frac{4\pi}{3N} \right )^{2/3}
\left ( \frac{r_{TF}}{l_0} \right )^4 \;  .
\ee

We solve numerically the system of seven equations (\ref{39}), (\ref{45}), (\ref{46}),  
(\ref{48}), (\ref{49}), (\ref{50}), and (\ref{54}) with the boundary conditions
$$
 n_0(1) = n_1(1) = \sgm(1) = 0 \;  .
$$
The chemical potential, shown in Fig. 1, is a monotonically increasing function of 
the gas parameter. The dimensionless sound velocity $s(r)$, as a function of the
dimensionless spatial variable $r$, is presented in Fig. 2 for different gas parameters.
The sound velocity increases for larger gas parameters. The spatial dependence of 
the dimensionless anomalous average $\sigma(r)$, for different gas parameters, 
is shown in Fig. 3. Note that the anomalous average is not monotonic with 
respect to $\gamma$, first increasing for $\gamma$ between $0$ and approximately 
$0.5$, but then decreasing. The spatial fraction of uncondensed atoms in Fig. 4
is maximal at the center, similarly to the total density. As a function of the gas 
parameter, it monotonically increases with respect to $\gamma$. 

\begin{figure}[ht!]
\centerline{
\includegraphics[width=7cm]{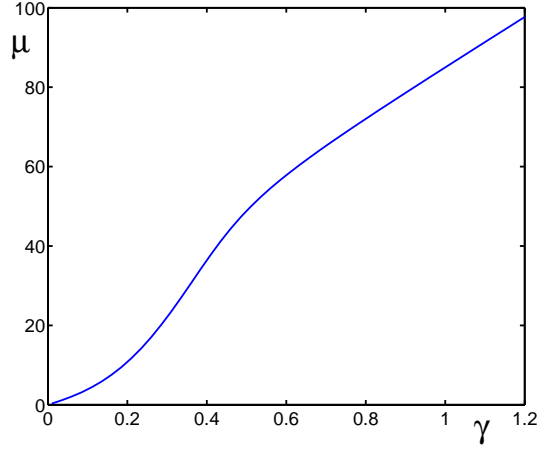} }
\caption{Dimensionless chemical potential $\mu$ as a function of the 
dimensionless gas parameter $\gamma$.
}
\label{fig:Fig.1}
\end{figure}

\begin{figure}[ht!]
\centerline{
\includegraphics[width=7cm]{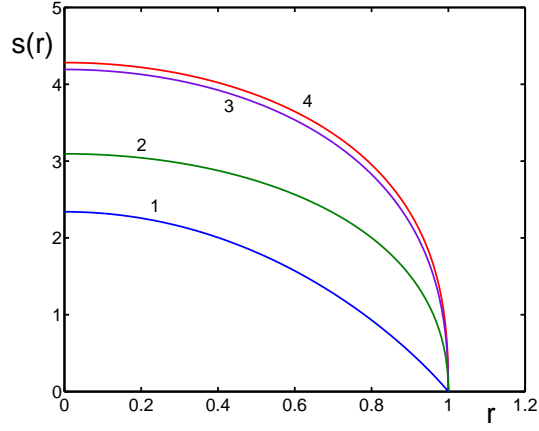} }
\caption{Dimensionless sound velocity $s(r)$, as a function of the spatial 
variable $r$, for different gas parameters: $\gm=0.1$ (line 1), $\gm=0.25$ (line 2),  
$\gm=0.5$ (line 3), and $\gm= 1$ (line 4). 
}
\label{fig:Fig.2}
\end{figure}

\begin{figure}[ht!]
\centerline{
\includegraphics[width=7cm]{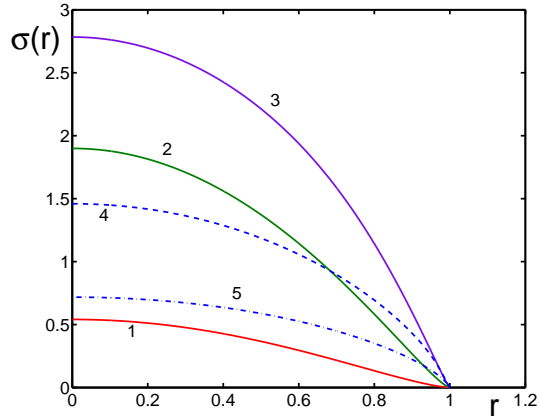} }
\caption{Dimensionless anomalous average $\sigma(r)$, as a function of the 
spatial variable $r$, for different gas parameters: $\gm=0.1$ (line 1), $\gm=0.25$ 
(line 2), $\gamma = 0.5$ (line 3), $\gamma = 1$ (line 4), and $\gamma = 2$ (line 5).
}
\label{fig:Fig.3}
\end{figure}

\begin{figure}[ht!]
\centerline{
\includegraphics[width=7cm]{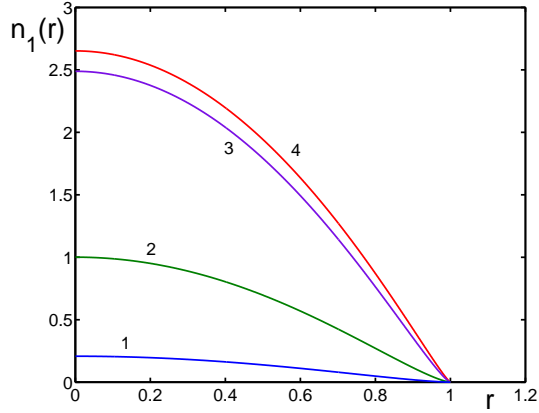} }
\caption{Spatial dependence of the fraction $n_1(r)$ of uncondensed atoms 
for different gas parameters: $\gamma = 0.1$ (line 1), $\gamma = 0.25$ (line 2), 
$\gamma = 0.5$ (line 3), and $\gamma = 1$ (line 4).
}
\label{fig:Fig.4}
\end{figure}

A very interesting is the behaviour of the condensate fraction $n_0$ as a function 
of the spatial variable. For small gas parameters $\gamma < 0.3$, the condensate  
fraction exhibits a maximum at the trap center, as in Fig. 5a. While, starting from 
$\gamma \approx 0.3$, the condensate fraction has a local minimum at the trap 
center, as in Fig. 5b. This trap-center depletion deepens with the increase of $\gamma$,
and the condensate is pushed out of the trap center to its boundary, as is demonstrated
in Fig. 6. Such a behaviour of the condensate fraction is in agreement with the
Monte Carlo simulations \cite{Dubois_5,Dubois_6}.

\begin{figure}[ht]
\centerline{\hbox{
\includegraphics[width=7cm]{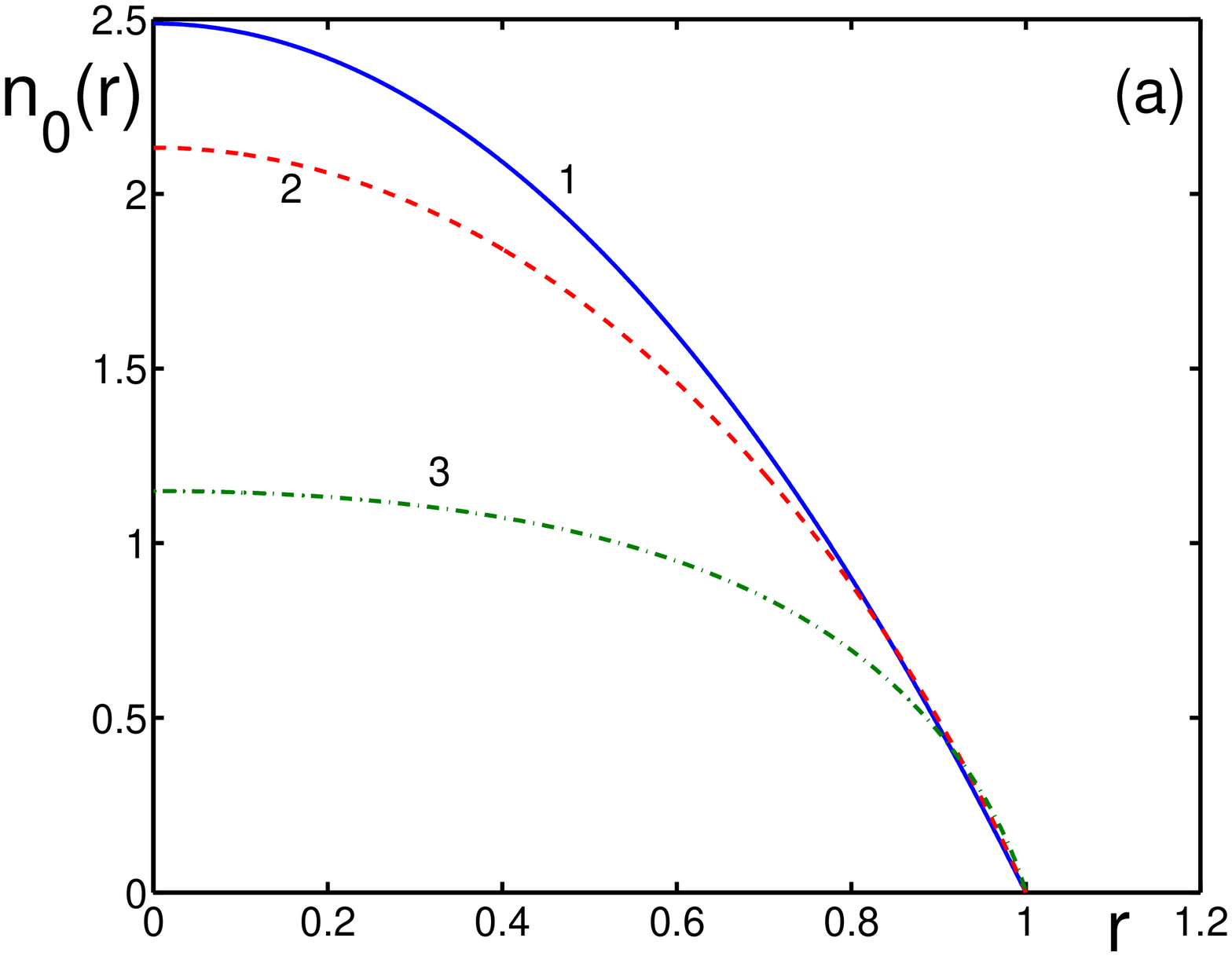} \hspace{2cm}
\includegraphics[width=7cm]{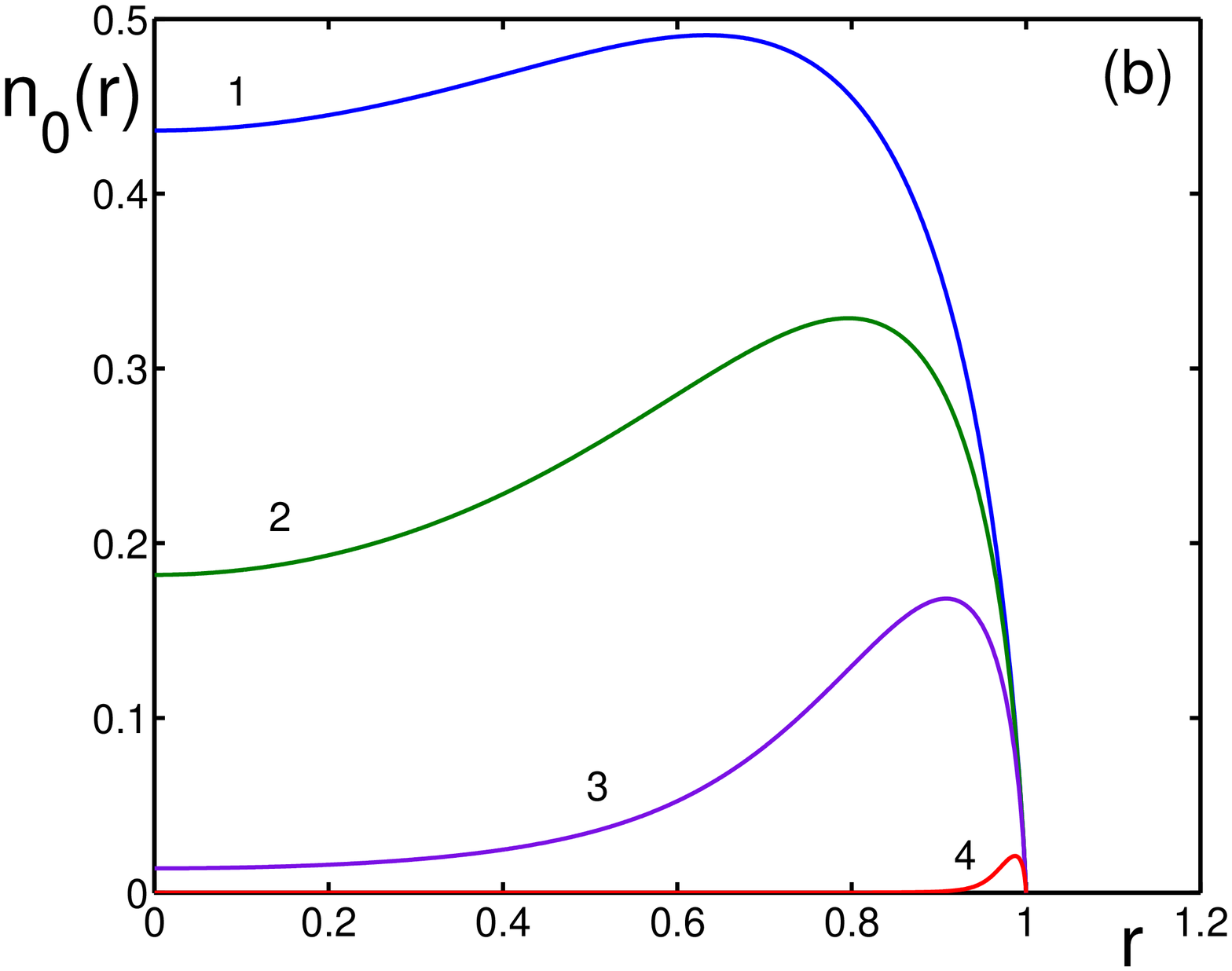} }   }
\caption{Spatial dependence of the condensate fraction $n_0(r)$ for different gas 
parameters: (a) no trap-center depletion for small $\gamma < 0.3$; here $\gamma = 0.01$ 
(line 1), $\gamma = 0.1$ (line 2), and $\gamma = 0.25$ (line 3); (b) trap-center depletion 
for $\gamma > 0.3$; here $\gamma = 0.35$ (line 1), $\gamma = 0.4$ (line 2), $\gamma = 0.5$
(line 3), and $\gamma = 1$ (line 4). 
}
\label{fig:Fig.5}
\end{figure}

\begin{figure}[ht!]
\centerline{
\includegraphics[width=7cm]{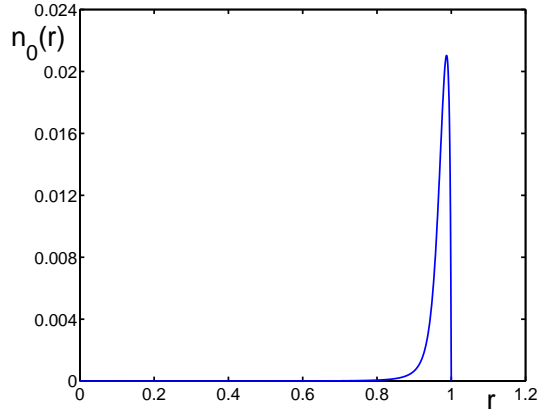} }
\caption{For $\gamma = 1$, the condensate fraction is pushed out of the trap center
and is located close to the boundary.
}
\label{fig:Fig.6}
\end{figure}
  
The effect of the trap-center depletion is due to the presence in the equation for the  
condensate fraction  (\ref{45}) of the negative terms, describing the fraction of 
uncondensed atoms, and the anomalous average. In the Bogolubov approximation,
when both these terms are absent, the appearance of the trap-center depletion is 
impossible, as is evident from equation (\ref{45}). This effect also does not occur, 
if one omits the anomalous average, while keeping the fraction of uncondensed atoms,
as is illustrated in Fig. 7. This emphasizes the necessity of accurately taking into 
account the anomalous average that in our case is given by equation (\ref{49}). 
Other forms of the anomalous average, which can be met in literature,  e.g. 
in \cite{Boudjemaa_46,Boudjemaa_47}, also do not lead to the trap-center depletion.

\begin{figure}[ht!]
\centerline{
\includegraphics[width=7cm]{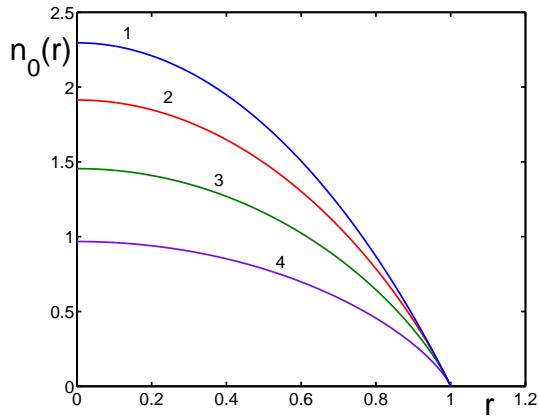} }
\caption{Role of the anomalous average in the effect of trap-center depletion. 
Omitting in equation (\ref{45}) the anomalous average results in the spatial behaviour 
of the condensate fraction with a maximum at the trap center for all gas parameters: 
$\gm=0.1$ (line 1), $\gm=0.25$ (line 2), $\gm=0.5$ (line 3), and $\gm=1$ (line 4). 
}
\label{fig:Fig.7}
\end{figure}

The mean condensate fraction in the trap, defined in equation (\ref{53}), is presented 
in Fig. 8, where it is compared with the results of the Bogolubov approximation (BA)
\cite{Javanainen_48} and of the diffusion Monte Carlo (DMC) analysis \cite{Dubois_6} 
listed in Table 1. The Bogolubov approximation essentially overestimates the condensate 
fraction for $\gamma < 0.5$, but becomes zero at $\gamma = 0.654$, where the DMC
simulations still show the condensate fraction of $10 \%$. The condensate fraction,
defined in equation (\ref{53}), is a bit lower than the DMC results. 

\begin{figure}[ht!]
\centerline{
\includegraphics[width=7cm]{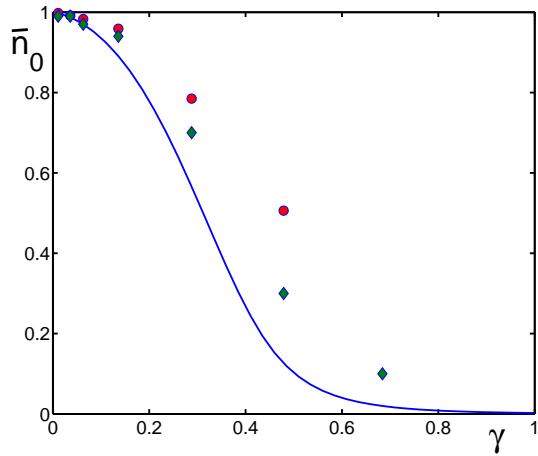} }
\caption{Mean condensate fraction in a trap as a function of the gas parameter 
(solid line), compared to the Bogolubov approximation (fat dots) and the diffusion
Monte Carlo simulation (diamonds).
}
\label{fig:Fig.8}
\end{figure}

\begin{table}[ht!]
\renewcommand{\arraystretch}{1.15}
\centering
\label{tab1}
\begin{tabular}{|c|c|c|} \hline
$\gm$   & $BA$ \cite{Javanainen_48}  &   $DMC$ \cite{Dubois_6}  \\ \hline
0.011   &  0.998                     &  0.99      \\ \hline
0.036   &  0.992                     &  0.99       \\ \hline
0.063   &  0.983                     &  0.97        \\ \hline
0.136   &  0.959                     &  0.94       \\ \hline
0.288   &  0.785                     &  0.7      \\ \hline
0.479   &  0.506                     &  0.3  \\ \hline
0.684   &   $-$                      &  0.1  \\ \hline
\end{tabular}
\vskip 2mm
\caption{Numerical results for the mean condensate fraction of the trapped Bose gas
in the Bogolubov approximation (BA) \cite{Javanainen_48} and in the diffusion Monte 
Carlo (DMC) analysis \cite{Dubois_6}. For $\gamma = 0.684$, the BA predicts a negative 
condensate fraction.}
\end{table}

Although a finite system, corresponding to a trapped Bose-condensed gas, is not the 
same as a homogeneous superfluid, it is interesting to compare the condensate fractions
in these two cases. The atoms of $^4$He at saturated vapor pressure are well 
represented \cite{Kalos_43} by hard spheres of diameter $a_s = 2.203$ \AA, which 
translates into the gas parameter $\gamma = 0.59$. The DMC simulations for the 
related $\gamma$ estimate the condensate fraction of $18 \%$. The recent experiments 
with bulk liquid helium \cite{Glyde_49,Diallo_50} give the zero temperature value 
$n_0 = (7.25\pm 0.75) \%$ at saturated vapor pressure and $n_0 = (2.8\pm 0.2) \%$ 
at the pressure close to solidification. In our case, for the gas parameter 
$\gamma = 0.59$, we find $n_0 = 4.4 \%$.

\section{Discussion}

We have analyzed the effect of the local condensate trap-center depletion, when for 
the gas parameter $\gamma > 0.3$ the spatial distribution of the condensate fraction 
exhibits a local minimum at the trap center, contrary to the maximum at the trap center 
for lower gas parameters $\gamma < 0.3$, as has been observed in Monte Carlo 
simulations \cite{Dubois_5,Dubois_6}. This effect cannot be described by solving the 
nonlinear Schr\"{o}dinger equation, either in the Thomas-Fermi approximation or in more 
refined approximations. 

But this effect can be explained employing the self-consistent approach to Bose-condensed 
systems 
\cite{Yukalov_9,Yukalov_15,Yukalov_16,Yukalov_17,Yukalov_20,Yukalov_33,Yukalov_34},
which is applied here for trapped atoms. In the frame of this approach, we use the Thomas-Fermi
approximation. Since the critical gas parameter, when the local-depletion effect develops, 
is $\gamma \approx 0.3$, one may ask whether the Thomas-Fermi approximation is applicable
in this range of $\gamma$.  To this end, we remember that the Thomas-Fermi approximation,
neglecting the kinetic energy term, is justified, when the interaction energy is much larger
than kinetic energy. The interaction energy per atom is of the order  
$E_{int} \sim \rho \Phi_0 \sim 4 \pi \rho a_s/ m$. And the effective kinetic energy of an atom 
is $E_{kin} \sim 1/2 m a^2 \sim \rho^{2/3}/ 2 m$, where $a$ is mean interatomic distance.
Therefore, $E_{int} / E_{kin} \sim 8 \pi \gamma$. The interaction energy is much larger than 
the kinetic energy, provided that $\gamma \gg 1/8 \pi =  0.0398$. The critical value of 
$\gamma = 0.3$ is an order larger than $0.04$, hence for the values of $\gamma$ around 
$0.3$ and higher, where the local depletion effect arises, the Thomas-Fermi approximation 
is certainly valid. 

The important role in the explanation of the local-depletion effect is played by the existence 
of the anomalous average, without which the effect does not occur. From the form of the 
anomalous average (\ref{13}) it follows that the modulus $|\sigma_1(\bf r)|$ of the anomalous 
average $\sigma_1(\bf r)$ defines the density of pair-correlated atoms \cite{Yukalov_9,Yukalov_20}. 
In that sense, the anomalous average is connected with pair correlations. The relation of the 
anomalous average to pair correlations can also be illustrated by considering the two-body 
scattering matrix \cite{Bogolubov_50} 
$$
S_k = \int \Phi(\br-\br') \; 
\lgl \; \vp_k^*(\br) \vp_{-k}^*(\br') \hat\psi(\br') \hat\psi(\br) \; \rgl \; d\br d\br'   
$$
describing the scattering of two particles from the medium to the momentum states 
$\varphi_k({\bf r})$. With the contact interaction potential (\ref{1}), the scattering matrix 
becomes 
$$
S_k = \Phi_0 \int |\;\vp_k(\br)\; |^2  \left [ \eta^2(\br) + \sgm_1(\br) \right ] \; d\br \; ,
$$
where the Bogolubov shift (\ref{3}) is used and the anomalous average is defined in 
equation (\ref{13}). Although the physical origin of the effect is rather clear, as being due 
to strong interatomic interactions pushing the condensate out of the trap center, but its 
theoretical description is quite delicate, requiring an accurate description of the 
anomalous average.

One more question that can arise when analyzing the results of the calculations, is whether
strong atomic interactions push uncondensed atoms from the trap center? To the first glance, 
looking at Fig. 4, one does not see a spreading of uncondensed atoms from the trap center 
under increasing interactions. However, one should not forget that in the figures we use the 
dimensionless units. While in dimensional form the atomic cloud radius depends on the 
interaction strength as 
$$
 r_{TF} = \sqrt{ \frac{2\mu_0}{m\om_0^2} } \; \approx \; 
\left ( 15 \; \frac{a_s}{l_0} \; N \right )^{1/5} \; l_0 \; .
$$
Therefore, under strengthening interactions, that is increasing the scattering length $a_s$, 
the cloud radius also increases. In that sense, atomic interactions do push atoms out of the 
trap center. Then the cloud shape becomes more flat. But the maximal density of uncondensed 
atoms remains at the trap center because normal atoms always prefer to gather at the location
of the minimal external potential, just spreading in space by enlarging the cloud size for reducing 
their density and interaction energy. Notice that employing dimensional quantities, we would 
have to fix two additional parameters, the trap oscillator frequency $\omega_0$ and the number 
of atoms $N$. While the scaling we have employed requires just the single gas parameter 
$\gamma$, which makes the consideration essentially more general.       

The found threshold gas parameter $\gamma = 0.3$ for the occurrence of the trap-center 
depletion of condensed atoms is in agreement with the Monte Carlo analysis \cite{Dubois_6}. 
The mean condensate fraction as a function of the gas parameter is close to the dependence 
found in Monte Carlo simulations.

\end{document}